\documentclass[%
 reprint,
 amsmath,amssymb,
 aps,
]{revtex4-2}

\usepackage{graphicx}
\usepackage{dcolumn}
\usepackage{bm}

\begin{document}

\preprint{APS/123-QED}

\title{Fraud detection in credit card transactions using Quantum-Assisted Restricted Boltzmann Machines}

\author{João M. Neto}
 \altaffiliation[e-mail:]{joao@niteqpuc.com.}
\author{Guilherme P. Temporão}%
\affiliation{%
 NITeQ, Electrical Engineering Department \\ Pontifícia Universidade Católica do Rio de Janeiro, Rio de Janeiro RJ 22451-900, Brazil
}%

\author{Gustavo C. Amaral}
\affiliation{
 Quantum Technology Department, The Netherlands Organization for Applied \\  Scientific Research, TNO, The Netherlands
}%

\begin{abstract}
Use cases for emerging quantum computing platforms become economically relevant as the efficiency of processing and availability of quantum computers increase. We assess the performance of Restricted Boltzmann Machines (RBM) assisted by quantum computing, running on real quantum hardware and simulators, using a real dataset containing 145 million transactions provided by Stone, a leading Brazilian fintech, for credit card fraud detection. The results suggest that the quantum-assisted RBM method is able to achieve superior performance in most figures of merit in comparison to classical approaches, even using current noisy quantum annealers. Our study paves the way for implementing quantum-assisted RBMs for general fault detection in financial systems.

\end{abstract}

\maketitle


\section{\label{sec:level1}Introduction:}

Quantum computing was proposed by Richard Feynman as a way to use quantum physics to effectively simulate quantum physics \cite{feynman2018simulating}. Since then, quantum computing has been evolving and quantum algorithms have been developed, such as Shor's algorithm \cite{shor1994algorithms}, capable of factoring large integers more efficiently than any supercomputer today.

Alongside algorithms, efforts have been made towards the physical implementation of quantum computers, such as a 3000 qubit ion trap computer \cite{chiu2025continuous}, photonic quantum computers \cite{zhong2020quantum}, and  superconducting quantum processors operating below the surface code error correction threshold \cite{google2025quantum}. There are several ways to physically implement quantum computing, such as gate-based quantum computing (GBQC) \cite{deutsch1985quantum}, adiabatic quantum computing (AQC) \cite{farhi2000quantum}, quantum computation based on anyons \cite{kitaev2003fault}, and quantum computation based on continuous variables \cite{braunstein2003quantum}. Since this work involves the resolution of different Quadratic Unconstrained Binary Optimization (QUBO) problems, an approximation of AQC, called quantum annealing (QA), was chosen. It has been shown to be a feasible way to solve said problems \cite{koshikawa2021benchmark, papalitsas2019qubo}.

AQC consists of initializing the qubits in a simple Hamiltonian, with a known ground state. This Hamiltonian is then adiabatically evolved to a more complex configuration, which will encode the target problem. The main idea is that, by adiabatically evolving the simple Hamiltonian while the qubits are initialized in its ground state configuration, the system will remain in its ground state. According to the adiabatic theorem \cite{kato1950adiabatic}, the final state of the qubits will correspond to the final state of the Hamiltonian, which corresponds
to the solution of the problem.

The aim of this study is to investigate the performance of machine learning algorithms (MLA) trained using quantum annealing and applied to classification problems. In particular, we combine a real world dataset, provided by the Brazilian fintech Stone, and compare the performances of classical and quantum-assisted-MLAs for fraud detection. A famous MLA is the Restricted Boltzmann Machine (RBM), an energy based network \cite{hinton2006fast} whose energy model is mathematically equivalent to QA \cite{adachi2015application}. By writing the energy function of the RBM as a QUBO problem, it is straightforward to transform it
into an Ising Hamiltonian, which can be effectively optimized by QA on D-Wave’s quantum computers and simulators \cite{ajagekar2020quantum}. This approach follows a similar framework presented in  \cite{ajagekar2020quantum}. Simultaneously, classical training is performed using Persistent Contrastive Divergence (PCD). To assist in the training procedure and the creation of RBMs, the Julia programming language module \texttt{QARBoM.jl} \cite{qarbom_2025} is deployed.

The paper is organized as follows. Section II provides a deeper dive into RBMs, QA, and Quantum Sampling (QS); section III focuses on the experimental implementation; section IV on the results; and section V on the final conclusions and future work directions

\section{Methods}

\subsection{Restricted Boltzmann Machines:}

A Restricted Boltzmann Machine (RBM) is a probabilistic generative machine learning model capable of handling supervised and unsupervised data, first introduced by Hinton et al \cite{hinton2006fast}. Since its training depends on an intractable partition function, approximations are needed to compute the parameter updates the procedure, as will be discussed in sec. \ref{training}. RBMs were developed as probabilistic generative models capable of learning and replicating the probability distribution of a dataset. Nevertheless, following \cite{larochelle2012learning}, they can also work as classifiers, known as Restricted Boltzmann Machine Classifiers (RBMCs).

\begin{figure}
    \centering
    \includegraphics[scale=0.5]{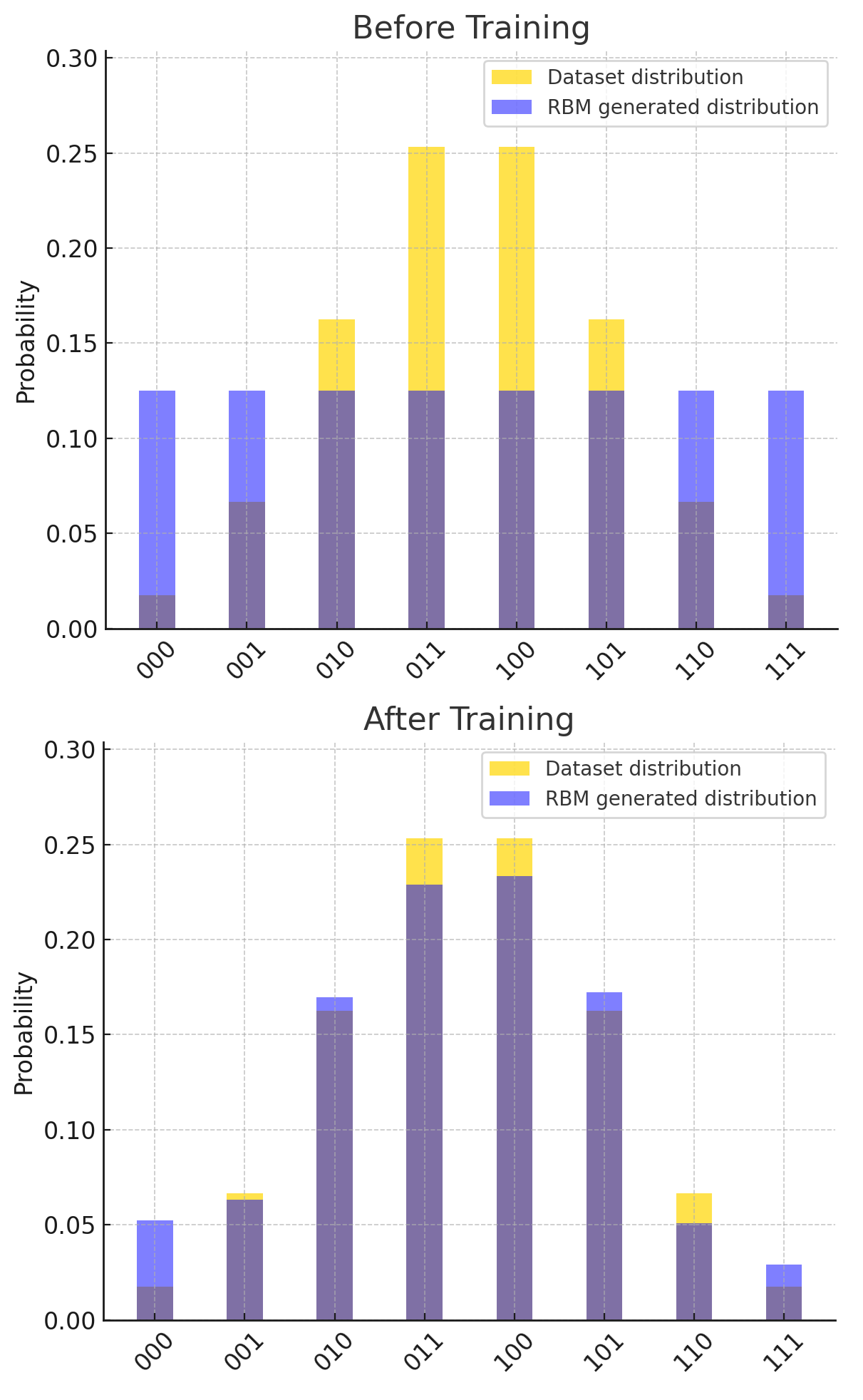}
    \caption{Representation of the probability distribution p(\textbf{v}) before (top) and after (bottom) training the RBM. Probability distribution of the dataset is shown in yellow, while the distribution of the data generated by the model can be seen in blue.}
    \label{fig:1}
\end{figure}

An RBM consists of two layers: a visible layer $(\textbf{v})$, where the visible units $\textbf{v}_i$ lie, and a hidden layer $(\textbf{h})$ with hidden units $\textbf{h}_i$. The main goal when training an RBM is the minimization of its energy function, which leads to the replication of the probability distribution of the dataset \cite{hinton2002training}. Adjusting the values of the biases and weights unlocks the possibility to adjust the probability distribution of the data generated by the model to match the probability distribution of the dataset, as pictorially displayed in Figure \ref{fig:1}.

The energy function of an RBM can be represented as follows \cite{hinton2006fast}: 
\begin{equation}
    E(\textbf{v}, \textbf{h}) = - \sum_{i,j} v_i W_{ij} h_j - \sum_i c_i v_i - \sum_j b_j h_j,
    \label{eq:1}
\end{equation}
where $c_i$ represents the bias terms for the visible units and $b_j$ the bias terms in relation to the visible units. With the energy function of Eq. \ref{eq:1}, we can define the joint probability distribution of \textbf{v} and \textbf{h} as follows:
\begin{equation}
        p(\mathbf{v}, \mathbf{h}) = \frac{1}{Z} e^{-E(\mathbf{v}, \mathbf{h})},
        \label{eq:2}
\end{equation}
where
\begin{equation}
        Z = \sum_v\sum_h e^{-E(\mathbf{v}, \mathbf{h})}
        \label{eq:Z}
\end{equation}
is a partition function. Although fundamental for the notion of probability, as it represents the normalization of the joint probability distribution, $Z$ depends on the summation of every single configuration of \textbf{v} and \textbf{h}. This makes the function calculation intractable, i.e., it can be unfeasible to directly compute the joint probability distribution for large problems in classical computers. 

The original RBM deals with binary data, making it incompatible with the characteristics of the dataset used in this work. An important extension of RBMs is the Gaussian Restricted Boltzmann Machine (GRBM) \cite{liao2022gaussian}, suitable for handling continuous-valued data. The marginal probability distribution of the visible layers is altered: instead of a Bernoulli distribution, we now have a continuous Gaussian distribution. The training methodology, discussed in the next Section, is analogous; the details that make it compatible with GRBMs will be highlighted when necessary.

\subsection{Training Restricted Boltzmann Machines:}
\label{training}

Training RBMs is a non-trivial task; recall that our goal is to minimize the energy function, which is equivalent to attributing high probability to examples generated akin to reality; in this case, the training dataset. In order to do so, one must first define a loss function to update the parameters. This loss function is generally defined as the log-likelihood of the probability distribution \cite{hinton2006fast}, which is given by:
\begin{equation}
    L(\textbf{V}) = -\sum_{i=0}^V \text{log}(p(\textbf{v}^{(i)})),
    \label{eq:5}
\end{equation}

By using stochastic gradient descent (SGD) \cite{robbins1951stochastic}, parameters of the RBM can be updated, making sure that the probability distribution of the data generated by the RBM is as close as it can get to the dataset. For the update of said parameters, the following useful mathematical relations are used \cite{hinton2012practical}:    
\begin{equation}
    \dfrac{\partial log(p(\textbf{v}^{(i)}))}{\partial \mathbf{W}} = \mathbf{h}_{\text{data}} \mathbf{v}_{\text{data}}^T - \mathbf{h}_{\text{model}} \mathbf{v}_{\text{model}}^T,
    \label{eq:7}
\end{equation}
\begin{equation}
    \dfrac{\partial log(p(\textbf{v}^{(i)}))}{\partial \mathbf{b}} = \mathbf{h}_{\text{data}} - \mathbf{h}_{\text{model}},
    \label{eq:8}
\end{equation}
\begin{equation}
    \dfrac{\partial log(p(\textbf{v}^{(i)}))}{\partial \mathbf{b}} = \mathbf{v}_{\text{data}} - \mathbf{v}_{\text{model}},
    \label{eq:9}
\end{equation}

The calculation of the gradient in relation to the positive phase in Eqs. \ref{eq:7}-\ref{eq:9} is trivial since the probability distribution of the data is given: the expected value of \textbf{$h_{data}$} can be estimated by sampling its conditional probability in relation to \textbf{v}. The negative phase, on the other hand, is not so trivial: for the calculation of the expected value, the joint probability of $\textbf{v}_{model}$ and $\textbf{h}_{model}$ would need to be sampled, which is intractable \cite{hinton2006fast}. Fortunately, one can circumvent this issue by calculating an approximation of the expected values using the {\it Persistence Contrastive Divergence} (PCD) method \cite{tieleman2008training} via the so-called -- classically inefficient -- Gibbs Sampling algorithm \cite{Geman1984}.

The training of a GRBM is done the same way as the RBM. The log-likelihood is the loss function, and the gradients can be updated the same way as before. The only difference is how the conditional probability distribution, necessary to calculate the gradients, is calculated \cite{Hinton2010}. By assuming that \textbf{v} is a continuous Gaussian distribution with zero mean and unitary standard deviation, we arrive at the following conditional probability distribution: 
\begin{equation}
    p(\mathbf{v} \mid \mathbf{h}) = \mathcal{N}(\boldsymbol{\textbf{c} + \textbf{W}^T\textbf{h}}, \mathbf{I}),
    \label{eq:16}
\end{equation}
where $\mathbf{I}$ represents the identity matrix. 

\subsection{Restricted Boltzmann Machines as a QUBO problem and Ising Model Formulation:}

Quadratic Unconstrained Binary Optimization (QUBO) problems \cite{punnen2022quadratic} aim at minimizing the following objective function:
\begin{equation}
     f(x) = \sum_i Q_{i,i} x_{i,i} + \sum_{i < j} Q_{ij} x_i x_j,  
\end{equation}
where $x$ represents binary variables and $Q_{i,j}$ are coefficients of the symmetric matrix \textbf{Q}, which determine linear and quadratic weights. With the objective function defined, we now need to write the energy function of the RBM as a QUBO problem, with slight modifications, as follows. First the visible, label, and hidden layers will all be concatenated in a single vector $\textbf{x} = (\textbf{v},\textbf{y},\textbf{h})$. Second, an ideal \textbf{Q} matrix must be found with the property that, when multiplied on the left by $\textbf{x}$ and on the right by $\textbf{x}^T$ it gives the correct loss function for Gaussian-Bernoulli RBMs \cite{hinton2012practical}. In summary, to construct the \textbf{Q} matrix, one should follow the following rules: 
\begin{enumerate}
    \item $Q_{i,i}$ represents the biases. 

    \item $Q_{i,j}, \text{represents the weights for } i \neq j $.

    \item $Q_{i,j} = 0, \text{ if } i \text{ and } j \text{ are the } \text{same type of unit.}$

    \item $Q_{i,j} = Q_{j,i} $.
\end{enumerate}

It is important to note that, when writing the energy function of the RBM as a QUBO problem, if the visible units are Gaussian, then a binary expansion is needed in order to still maintain the overall structure of the problem. This increases the number of variables and will eventually become a bottleneck when implementing the RBM in a quantum computer, which is the focus of this work. Although an issue, it also represents a necessary step in order to achieve the goal of assisting RBM training with a quantum computer.    

With the \textbf{Q} and \textbf{x}, one can finally write the RBM as a QUBO problem. By making used of a simple transformation, presented below, it is possible to convert the QUBO to an Ising Hamiltonian \cite{ajagekar2020quantum}. This enables one to perform the optimization efficiently when assisted by an adiabatic quantum computer such as D-Wave's system \cite{ajagekar2020quantum}. 
\begin{equation}
        H_{\text{Ising}} = C + \sum_i h_i s_i + \sum_{i<j} J_{ij} s_i s_j,
\end{equation}
where $C$, $h_{i}$, and $J_{i,j}$ are constants that will define the end goal of the QA process. 

\subsection{Quantum-Assisted Restricted Boltzmann Machines:}

With the QUBO model in hand and the conversion to the Ising Model determined, it is important to identify where the a quantum annealer system can make the training of RBMs more efficient. To that end, let us recall the total Hamiltonian of the system, and how its adiabatic evolution. The generic formulation of the Hamiltonian in QA systems is given by: 
\begin{equation}
    H(s) = a(s)H_D + b(s)H_P,
    \label{eq:24}
\end{equation}
where $s$ is a normalized time parameter, $H_D$ and $H_P$ are the driver and the problem Hamiltonians. $a(s)$ and $b(s)$ are two functions with the following pattern: when s $\approx$ 0, then $a(s) \approx 1$ and $b(s) \approx 0$; when s $\approx$ 1, then $a(s) \approx 0$ and $b(s) \approx 1$. Essentially, the process consists of a gradual transition from the driver Hamiltonian to the problem Hamiltonian. 



Ideally, we would encode the target problem Hamiltonian as the QUBO problem that minimizes the energy function of the RBM, and subsequently solve it via the QA. This, however, is not possible due to NISQ quantum computers \cite{preskill2018quantum} being prone to noise. Be that as it may, it is still possible to optimize the model and update its parameters using PCD, while avoiding the inefficient Gibbs Sampling algorithm and replacing it with the QA. One important aspect to understand is that there is a non-zero probability that the QA process will not occur adiabatically when running in real hardware. In other words, the system has a chance to escape the ground state and evolve towards an excited state, following the Boltzmann Distribution:
\begin{equation}
    P(x) = \frac{1}{Z} e^{-\beta H_P(x)},
    \label{eq:26}
\end{equation}
where $\beta$ is the inverse temperature parameter, linking the energy of a configuration to its sampling probability. In order to circumvent the issue, one can call the QA routine $N$ times; with sufficient samples, one can statistically estimate the values of $\textbf{v}_{model},y_{model}$ and $\textbf{h}_{model}$ using quantum hardware, without explicit computation of the intractable partition function. This process is known as Quantum Sampling (QS).

\section{Implementations}
\subsection{Stone's Dataset:}

The dataset for the project was provided by Brazilian fintech \textbf{Stone}, and the study revolves around identifying fraudulent transactions along with many non-fraudulent ones. The dataset covers a time span of three months with transaction data from all over Brazil. The information is anonymized in accordance with Brazilian data protection laws, so neither user information nor balance amount is disclosed. In total, there are 145 million transactions, out of which 19 thousand are fraudulent, i.e., a highly unbalanced dataset. RBM training generally suffers from such a feature, which is why dataset balancing procedures are further discussed in Section IV.B.

Furthermore, data pre-processing methodologies were introduced while consulting with Stone's anti-fraud team, making sure not to introduce unwanted bias to the dataset. These included, for the categorical data, label encoding, binary encoding, hashing, and ordinary encoding, with the most effective being the so-called One-Hot Encoding \cite{ruppert2004elements}. Numerical data was treated so that its distribution would follow the guidelines of a GRBM; on other words, entries were normalized leading to a respective Gaussian distribution with zero mean and unitary standard deviation. This can be done by applying the Z-Score function to all numerical columns:
\begin{equation}
    Z = \frac{x-\mu}{\sigma},
    \label{eq:27}
\end{equation}
where $x$ represents the numerical data and $\mu$ and $\sigma$ represent the mean and the standard deviation of the column, respectively.

Finally, dataset filtering was implemented with the goal of restricting the size of the problem, an important procedure in order to enable embedding into real QA hardware. To determine the final filtered dataset, the correlation matrix of the instance features was calculated and those with the highest correlations were eliminated. Furthermore, relevance of the remaining instance features was studied using the so-called CatBoost model \cite{hancock2020catboost} and eliminating those with the least importance.

\subsection{D-Wave's Functionality and Embedding:}

D-Wave's Hamiltonian is given by:
 \begin{align}
 \begin{split}
 \mathcal{H}_{\text{Ising}} &= 
 \frac{A(s)}{2} \left( \sum_i \hat{\sigma}_x^{(i)} \right)\\
 &+ \frac{B(s)}{2} \left(\sum_i h_i \hat{\sigma}_z^{(i)} + \sum_{i > j} J_{ij} \hat{\sigma}_z^{(i)} \hat{\sigma}_z^{(j)},
 \right)
 \end{split}
 \end{align}
where $A(s)$ is defined as the transverse energy and $B(s)$ is known as the energy applied to the problem Hamiltonian. Their functionality is essentially the same as in eq. \ref{eq:24}: for $s \approx 0$ then $A(s) >> B(s)$ and for $s \approx 1$ then $B(s) >> A(s)$.

\begin{figure}
    \centering
    \includegraphics[width=\linewidth]{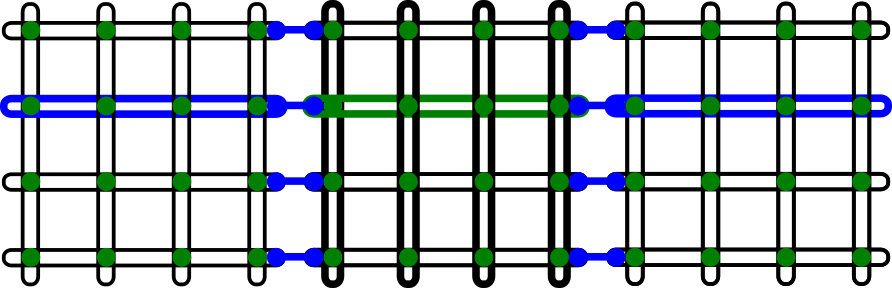}
    \caption{Representation of a 3x3 Chimera grid, a simplified model of the D-Wave 2000Q quantum annealer. The loops represent the qubits, the blue dots represent the external couplings (which allow for the conection of two unit cells), and the internal couplings are represented by green circles (allowing the conection of one or more qubits inside the unitary cell).
    Image source: D-Wave's Documentation}
    \label{fig:3}
\end{figure}

D-Wave's computers consist of three different topologies: Chimera, Pegasus, and Zephyr. They all work following the same principles, but can achieve different connectivity between physical qubits. The structure is composed of unit cells, a block of interconnected qubits that repeats itself, forming a grid. Different unit cells are connected to each other via couplers, allowing qubits to interact with neighboring qubits and, consequently, neighboring cells. A simplified representation of grids, cells, and qubits is depicted in Fig. \ref{fig:3}.

For an ideal problem embedding, each hidden and visible unit of an RBM would be mapped to a single qubit. The word 'ideal' is used here since, in order to perform QA, D-Wave's systems need to form a closed loop within a unit cell \cite{choi2008minor}. That is not always possible, and when it isn't, one needs to make use of a process called {\it minor embedding}, which consists of chaining various qubits together to represent a more robust logical qubit. When dealing with high dimensional data, one runs into the limitation of the number of qubits in the system. This made the minor embedding process the primary bottleneck, which was overcome during the research project by restricting the dataset features as discussed in the previous Section.  

\subsection{Simulated Annealing:}

An intermediate step between the strictly classical training procedure to the quantum-assisted one is the use of Simulated Annealing (SA), a feature offered by D-Wave's simulators \cite{choi2008minor}. Although this method does not take advantage of the full features of quantum annealing, its implementation is more accessible: minor embedding becomes less restrictive and the size of the RBM (number of visible and hidden units) can reach higher values than for the real hardware implementation.

D-Wave implements SA by discretizing time in different steps called {\it sweeps}. In each sweep, the individual qubits are updated in accordance with a transition probability derived from a Path Integral Monte Carlo (PIMC) estimation \cite{ceperley1995path} in order to simulate and approximate the effects of quantum fluctuations. At the end of the process, the system should approximate a Boltzmann distribution as described in eq. \ref{eq:26}. 

\subsection{QARBoM.jl:}

In order to help with the creation, training, and optimization of hiperparameters of the RBMs, Ripper et al. \cite{qarbom_2025} created \texttt{QARBoM.jl}, an open-source Julia programming language package that provides a one-stop shop of comparative training of RBMs. It provides an easy and useful way to create and train RBMs, whether using strictly classical methods or quantum-assisted ones. \texttt{QARBoM.jl} also provides a very robust backbone when dealing with QUBO problems because of its integration with the \textit{QUBO.jl} package \cite{xavier2023qubo}, which converts various optimization problems to QUBO. \texttt{QARBoM.jl} also directly integrates with yet another Julia package, \texttt{DWave.jl} \cite{xavier2023qubo}, providing seamless integration with D-Wave's systems.

\section{Experiments}

This Section presents the comparative analysis between different methodologies for training RBMCs towards fraud detection based on experimental results. The three chosen methodologies include: strictly classical PCD; simulated-annealing-assisted PCD; quantum-annealing-assisted PCD. The selected figures of merit for comparison are Accuracy, Precision, Recall and F1-Score, which are standard metrics for classifiers \cite{powers2020evaluation}. It should be mentioned that the testing procedure consisted of using the same database described in Section III-A for all three cases, always with 50 epochs of training.

\subsection{Dataset Balancing:}

As previously discussed, the dataset is highly unbalanced, to which RBMCs are inherently sensitive. Consequently, four complementary strategies were employed to alleviate the adverse impact of this imbalance: oversampling, undersampling, oversampling followed by undersampling (and vice versa), and manual undersampling.

Oversampling consists on the usage of the SMOTE algorithm \cite{chawla2002smote}. Unfortunately, it provided less than ideal results. Since the minority class was significantly small, this caused the model to learn predominantly from synthetic rather than real samples, thus producing an undesirable effect, leading to frequent misclassifications. Undersampling consists of implementing the well-known Tomek Links algorithm \cite{tomek1976two}. As with oversampling, results were far from ideal, as the minority class was extremely small. This caused the dataset to be reduced to a few hundred lines, rendering it nearly useless, as there was little to no data left for the model. Oversampling followed by Undersampling suffered from the same aforementioned limitations. The minority class was simply too small to allow for a reconstruction that enabled the model to be trained primarily on real data. Reversing the order of the approaches was also tested but led to the same issue. 

Manual Undersampling was ultimately required due to the length of the dataset and a straightforward, but effective, procedure was implemented: given all the fraudulent instances, the same number of non-fraudulent ones was randomly selected and combined. This resulted in a dataset of 39 thousand transactions with a 50-50 (unbiased) split. Manual Undersampling exhibited the best training features since it only uses non-synthetic data: the model can learn to detect fraudulent transactions in a more general environment. For the evaluation of the classification metrics, the dataset was divided into training and test sets, with a 80\%-20\% split, respectively. All the results shown in the later sections of this chapter are metrics in relation to the test set.

\subsection{Hyperparameter Search:}

\begin{figure*}[t]
    \centering
    \includegraphics[width=0.9\linewidth]{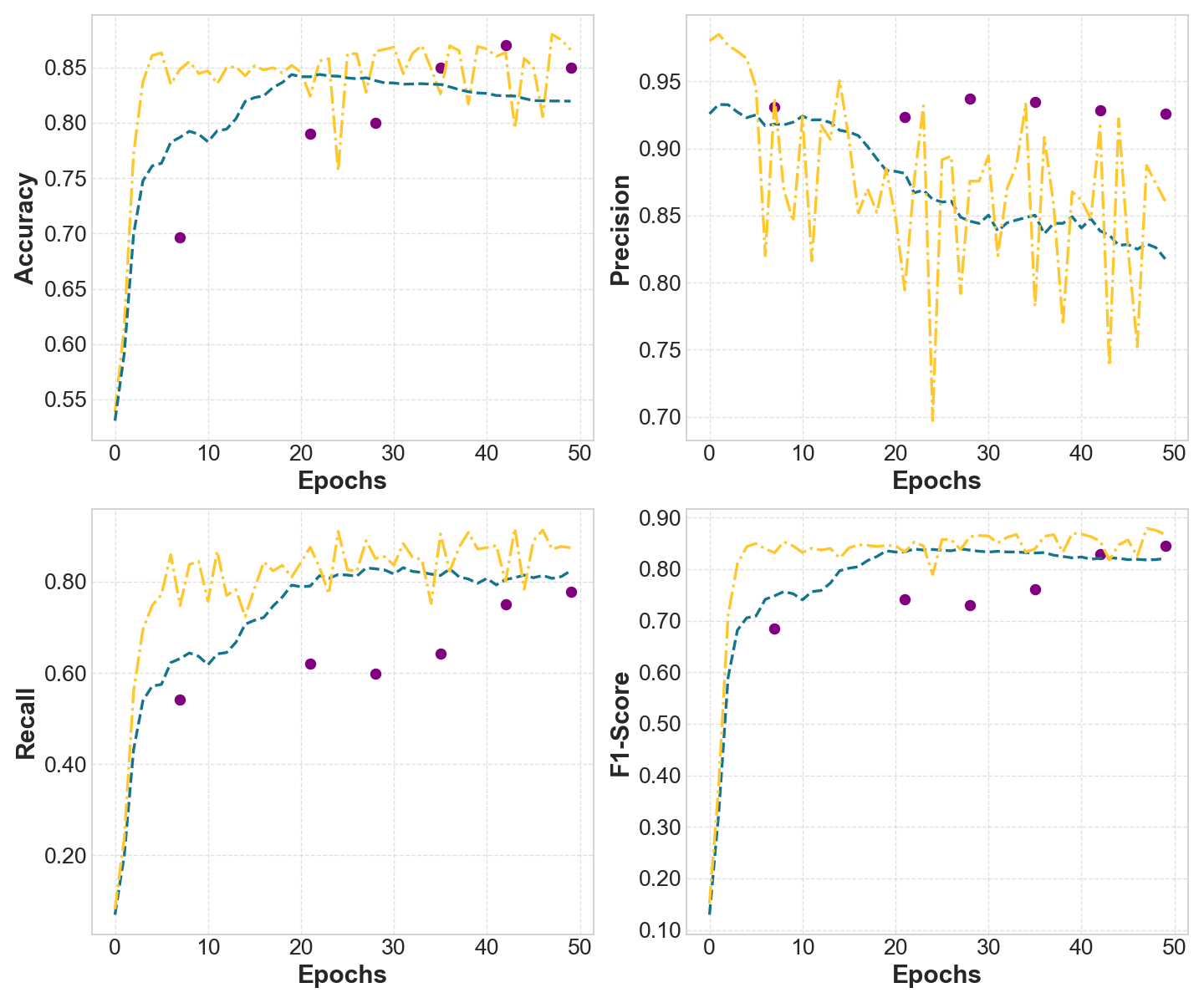}
    \caption{Comparison between the three training methods: strictly-classical PCD (blue), simulated-annealing-assisted training (yellow), and quantum-annealing-assisted training (Purple). Training was executed in 50 epochs; for each epoch, \textit{Accuracy}, \textit{Precision}, \textit{Recall}, and \textit{F1-Score} were calculated as comparative metrics. Due to limited access to D-Wave's system, only certain values were available in the quantum-annealing-assisted training; they are plotted as purple dots in the figure. The optimum configuration achieved by each training method corresponds to epoch 25, 34, and 49, respectively.}
    \label{fig:5}
\end{figure*}

Hyperparameter search consisted of finding the best possible combination of hyperparameters for the three tested training methodologies. The first hyperparameter analyzed was the number of hidden units in each RBMC. An excessively large number of units can lead the model to overfit, while too few units may result in underfitting. The second hyperparameter examined was the learning rate, considering its initial and final values as well as the decay schedule that governs its reduction throughout training. An extensive grid search was conducted to optimize these hyperparameters, aiming to identify the configuration that maximized both recall and precision. The best hyperparameter combinations found for each RBM can be found in Figure \ref{fig:5}.

For the strictly classical training methodology, results revealed that 200 hidden units were ideal, with an exponentially decreasing learning rate reaching zero by the end of the training procedure. For simulated-annealing-assisted training, results revealed that 65 hidden units were ideal with the same learning rate as for the classical one. For quantum-annealing-assisted, the best number of hidden units was the same as for the simulated case: 65. The learning rate, on the other hand, differed significantly: instead of an exponential decay from an initial value towards zero, the learning rate was defined as a smooth and controllable exponential decay function, ensuring gradual convergence while avoiding premature stagnation:
\begin{equation}
\text{LR}(\eta_0, \lambda, \eta_f) = \eta_0 \cdot \exp(-\lambda \cdot (t - 1)) + \eta_f,
\end{equation}
where $\eta_0$ + $\eta_f$ represents the starting value, $\eta_f$ the final value of the decay, $\lambda$ the speed in which the decay occurs, and, finally, the current epoch $t$.

\subsection{Results:}

A noteworthy distinction between strictly-classical and -- either simulated or real-hardware -- quantum-assisted training arises with respect to batch size. In classical training, larger batch sizes are typically beneficial: in our experiments, batches of 512 samples were used. These batches are fed into the training model instead of the entire dataset and translate to heavier or lighter training sessions \cite{Hinton2010}. For the different flavors of quantum-assisted training, however, the opposite tends to hold, with smaller batch sizes yielding superior performance.

Simulated-annealing-assisted training results were obtained by training RBMs optimized according to the aforementioned hyperparameter search methods and using D-Wave’s simulators. As with the classical model, the initial results were not promising, indicating the need for dedicated hyperparameter tuning tailored specifically for this procedure. Metric evolution throughout different epochs exhibit significant oscillations, which is likely associated to the limited number of sweeps during the simulation. A larger number of sweeps translates into impractical run times, exceeding hours per epoch, and had to be constrained during the comparative analysis.

Quantum-annealing-assisted training results were obtained by training RBMs using D-Wave’s quantum processors, following the same hyperparameter search optimization procedure described above. Surprisingly, both learning rates had to be significantly reduced compared to those used for the simulated-annealing-assisted training; using the same learning rates as in the previous methods caused the model to diverge. This behavior may stem from the fact that parameter updates derived from quantum-generated samples exhibit different statistical properties from those obtained through simulated sampling.

Real quantum sampling -- and all quantum annealing implementations used in this project -- was made possible through the Leap Program \cite{dwave_cloud_platform}, which provides cloud-based access to D-Wave’s quantum processors. Quantum annealing took approximately 1 hour and 50 minutes to complete the 50-epoch routine. The final training configurations are extracted from the highest ranking epoch with respect to the F1-score metric (the most relevant to the fraud detection problem according to Stone's anti-fraud team). The results are summarized in Table 1.

\begin{table}[]
\begin{tabular}{|c|c|c|l|l|l|}
\hline
Method & \multicolumn{1}{l|}{Accuracy} & \multicolumn{1}{l|}{Precision} & Recall & F1-Score & Total Time \\ \hline
PCD    & 84\%                          & 86\%                           & 81\%   & 83\%     & 2 minutes  \\ \hline
SAS    & 85\%                          & 94\%                           & 82\%   & 88\%     & 4 hours    \\ \hline
QS     & 85\%                          & 91\%                           & 79\%   & 84\%     & 1.8 hours  \\ \hline
\end{tabular}
\label{tab:1}
\caption{Comparison of performance metrics at the optimum training epoch for each model, evaluated over a 50-epoch training process.}
\end{table}

\subsection{Comparison:}

Fig. \ref{fig:5} presents the four different metrics evolution throughout the epochs for the three compared methodologies. As one can see, all models were able to achieve similar metrics, with simulated-annealing-assisted training reaching slightly higher values across all metrics, even considering its less consistent evolution (due to the limited number of sweeps, as discussed before). It is noteworthy that, even though the simulated- as well as quantum-annealing-assisted training methodologies produced similar performances with respect to the strictly classic one while running on a much lighter configuration. This is evidenced by the 55 visible, 200 hidden, 512 samples of the latter and 55 visible, 65 hidden, and 32 samples of the former two. This comparative result evidences the advantage introduced even by NISQ era platforms.

Another important figure of merit for comparison is training time, particularly given the substantial discrepancy observed across the three approaches. The fastest model, by a wide margin, was the classical one, requiring only two to three minutes to complete training. The slowest model, without question, was the simulated-annealing-assisted training method, taking approximately four hours to train. This difference in runtime can be attributed to the combination of a small batch size and the use of a Windows 11 machine. In particular, Python calls invoked from Julia incur substantial overhead on this operating system, significantly increasing execution time. A silver lining, however, is that despite being the slowest method overall, it reaches optimal performance relatively quickly, suggesting that a 50-epoch training routine was largely excessive for this case. Quantum Sampling was neither as fast as the strictly classical method nor as slow as simulated annealing one, but its training time still fell on the slower end of the spectrum.

\section{Conclusion}

In this study, a detailed investigation was carried out comparing the performance of different RBM models, including classically trained RBMs and quantum-assisted RBM models trained using both simulated and quantum annealing on real hardware. The results indicate that, for this specific problem, methods that deviate from the strictly classical methods already exhibit better performance, which can be extrapolated for more robust platforms than the current NISQ era ones. We verified that performance and runtime characteristics offer a compromise relationship: higher performance was associated with longer runtime and vice-versa. Whether this advantage is universal or specific to the characteristics of this dataset remains an open question and requires further study before a definitive conclusion can be drawn. Future work will explore new approaches to quantum-assisted machine learning, such as the quantum kernel method \cite{rebentrost2013quantum} for support vector machine training, as well as methods to further enhance the current execution of quantum-assisted training in, e.g., gate-based platforms using the Quantum Approximate Optimization Algorithm (QAOA) framework \cite{diez2023quantum}.

\begin{acknowledgments}
This work was supported by Stone Institui{\c c}{\~ a}o de Pagamento SA (funding number 2551.00.00) and CNPq (grant number 409596/2022-1). The group would like to extend its sincere gratitude to M. Ruas, from the Stone Anti-Fraud Team, whose assistance was instrumental to the completion of this work.

We also wish to express our most heartfelt thanks to the members of the Interdisciplinary Nucleus for Quantum Technologies (NITeQ) team of PUC-Rio, A. Wanick, B. Wolf, C. Nascimento, P. Ripper and M. dos Magos, for their unwavering support and fruitful discussions throughout this project.
\end{acknowledgments}

\bibliography{apssamp}

\end{document}